%% file: MoS2_NL_temp_v2.tex
\documentclass[journal = nalefd, manuscript = letter, layout =twocolumn]{achemso}

\usepackage[T1]{fontenc}
\usepackage{amsfonts}
\usepackage{amssymb}
\usepackage{amsmath}
\usepackage{xcolor}
\usepackage{verbatim}
\usepackage[percent]{overpic}
\usepackage{soul}

\SectionNumbersOn

\input{ourmacros}

\author{Enrico Perfetto}
\affiliation{Dipartimento di Fisica, Universit\`{a} di Roma Tor Vergata,
Via della Ricerca Scientifica 1, 00133 Rome, Italy}
\alsoaffiliation{INFN, Sezione di Roma Tor Vergata, Via della Ricerca Scientifica 1, 00133 Rome, Italy}
\email{enrico.perfetto@roma2.infn.it}

\author{Gianluca Stefanucci}
\affiliation{Dipartimento di Fisica, Universit\`{a} di Roma Tor Vergata,
Via della Ricerca Scientifica 1, 00133 Rome, Italy}
\alsoaffiliation{INFN, Sezione di Roma Tor Vergata, Via della Ricerca Scientifica 1, 00133 Rome, Italy}

\title{Real-time $GW$-Ehrenfest-Fan-Migdal method for nonequilibrium 2D 
materials}

\begin{document}


\begin{abstract}
Quantum simulations of photoexcited low-dimensional systems
are pivotal for understanding how to functionalize and integrate 
novel two-dimensional (2D) materials in next-generation optoelectronic devices.
First principles predictions are extremely challenging 
due to the simultaneous interplay of 
light-matter, electron-electron and electron-nuclear 
interactions. We here present an 
advanced ab initio many-body method which
accounts for quantum coherence and non-Markovian effects while 
treating electrons and nuclei on equal footing, thereby preserving 
fundamental conservation laws like the total energy. 
The impact of this advancement is 
demonstrated through real-time simulations of the complex multivalley dynamics
in a molybdenum disulfide 
(MoS$_{2}$) monolayer pumped above gap. Within a single framework 
we provide a parameter-free description of the coherent-to-incoherent 
crossover, elucidating the role of microscopic and collective 
excitations in the dephasing and thermalization processes.
\end{abstract}

\maketitle

\section{Introduction}

The ultrafast electronic, transport and optical properties 
of semiconductors are determined
by their  response to a photoexcitation. Understanding 
the underlying microscopic  dynamics is crucial for technological applications in
optoelectronics~\cite{opto4,opto1,opto2,opto3}, 
photovoltaics~\cite{photov1,photov2,photov3} and 
photocatalysis~\cite{cata1,cata2,cata3}.
The number of experimental time-resolved studies in 
advanced materials like  transition metal dichalcogenides (TMD)
has virtually exploded over the
last ten years~\cite{Mak2016,Tan6b00558,ceballos2017,Maiuri9b10533}.
The emerging universal picture is that the quantum dynamics initiated 
by an ultrafast  
excitation is characterized by two  distinct and consecutive regimes: 
an initially coherent regime featuring a macroscopic polarization 
that oscillates on the femtoseconds time-scale and a subsequent 
incoherent (or depolarized) regime during which the excited carriers 
 bring to completion the thermalization and then cool down~\cite{Lloyd_Hughes_2021}.
A unified theoretical framework capable of simulating the real-time 
dynamics from the coherent to the incoherent regime would be utmost 
desirable for parameter-free predictions. 

The challenge is to simultaneously account for material-specific 
ab initio electronic and phononic band structures, light-matter, 
electron-electron ({\it el-el}) and electron-phonon ({\it el-ph}) 
interactions. Methods like Time-Dependent Density-Functional 
Theory~\cite{doi:10.1063/1.1904586,tddftbook2012,doi:10.1063/1.4953039} in the typical adiabatic flavour~\cite{doi:10.1063/1.5142502,PhysRevB.85.045134} 
or time-dependent $GW$ Bethe-Salpeter~\cite{PhysRevB.84.245110,jiang_real_2021}
account for  both polarization and carrier
dynamics but are limited to the coherent kinetics as correlation-induced dephasing mechanisms
(responsible for the damping of the polarization) and relaxation 
processes (responsible for thermalization and cooling) are missing. 
Approaches based on the Boltzmann Equation (BE)\cite{kadanoff1962quantum,jauho-book}
treat relaxation processes at a Markovian level and are therefore suited 
to simulate thermalization and cooling  (incoherent regime). 
In fact, BE has no acces to the coherent dynamics since 
the polarization is assumed to vanish from the outset.
The microscopic description of the coherent-to-incoherent
crossover has therefore remained elusive so far. 
In the coherent regime scattering processes are intrinsically 
time-nonlocal and it is precisely 
such non-locality that is necessary to capture the 
correlation-induced dephasing~\cite{jauho-book}. State-of-the-art methods based on the 
more advanced semiconductor Bloch equations~\cite{SB1,HaugKochbook} treat the polarization 
degrees of-freedom semiclassically and ignore their 
feedback on the carrier dynamics~\cite{Marini2013,Steinhoff2016,MolinaSanchez2017}, thus violating the conservation of 
the total energy. Another difficult aspect of the crossover is that 
for laser pulses shorter than the screening buildup time 
the effective interaction between hot carriers 
is neither the bare Coulomb interaction nor the 
fully developed screened one~\cite{HaugKochbook,PhysRevB.62.7179}. 
Addressing all this physics calls for
a non-Markovian treatment of electronic and 
phononic correlations within a framework that handles  
polarization and carrier degrees-of-freedom on equal footing.
In this work we provide a versatile ab initio method based 
on nonequilibrium  Green's function (NEGF)~\cite{svl-book} which encompasses all 
required features.

We build on a recent time-linear scheme~\cite{PhysRevLett.124.076601,PhysRevB.104.035124,PhysRevB.105.125134,GWtilde} based
on the Generalized Kadanoff-Baym Ansatz
for electrons~\cite{PhysRevB.34.6933} and phonons~\cite{PhysRevLett.127.036402} 
to solve the NEGF equations
for the nonequilibrium electronic {\em and} phononic density matrices.
We pioneer the simultaneous inclusion of  {\it el-el} correlations at the $GW$ level~\cite{PhysRevLett.128.016801} and 
{\it el-ph} correlations at the Ehrenfest plus Fan-Migdal (FM) 
level~\cite{PhysRevLett.127.036402}.
No further approximations like time-local screening or frozen phonons 
are introduced; the polarization decays  without any 
phenomenological or semi-empirical dephasing rate. 
Simulations up to hundreds of femtoseconds allow us 
to ``watch'' the photoexcitation event and the subsequent coherent 
dynamics, the coherent-to-incoherent crossover and the slow trend 
towards the thermalization. 
We stress that excitonic effects are fully included 
in the dynamics. 
In fact, real-time $GW$ with a {\em statically} 
screened interaction $W$ is equivalent (in linear response) to solving 
the Bethe-Salpeter equation~\cite{PhysRevB.84.245110,jiang_real_2021}. Our approach does not rely on 
effective Hamiltonians written in terms of composite bosons describing 
excitons; free carriers and excitons both participate to the electronic 
density matrix. We improve over current real-time $GW$ methods in three ways:
(i) we use a {\em dynamically} screened interaction $W$~\cite{PhysRevLett.128.016801}; (ii) we account
for the interaction with both coherent (Ehrenfest) and incoherent 
(Fan-Migdal) phonons and (iii) we conserve the total energy of the el-ph system~\cite{PhysRevB.105.125134}.
The evolution of the density matrix does therefore embraces the build-up of
screening, retardation effects, formation of coherent excitons,
and decay into incoherent exciton-polarons (low density)~\cite{PhysRevB.103.245103}
or the coherent exciton Mott transition (high 
density)~\cite{PhysRevLett.128.016801}.

\begin{figure*}[tbp]
\centerline{\includegraphics[width=.5\textwidth]{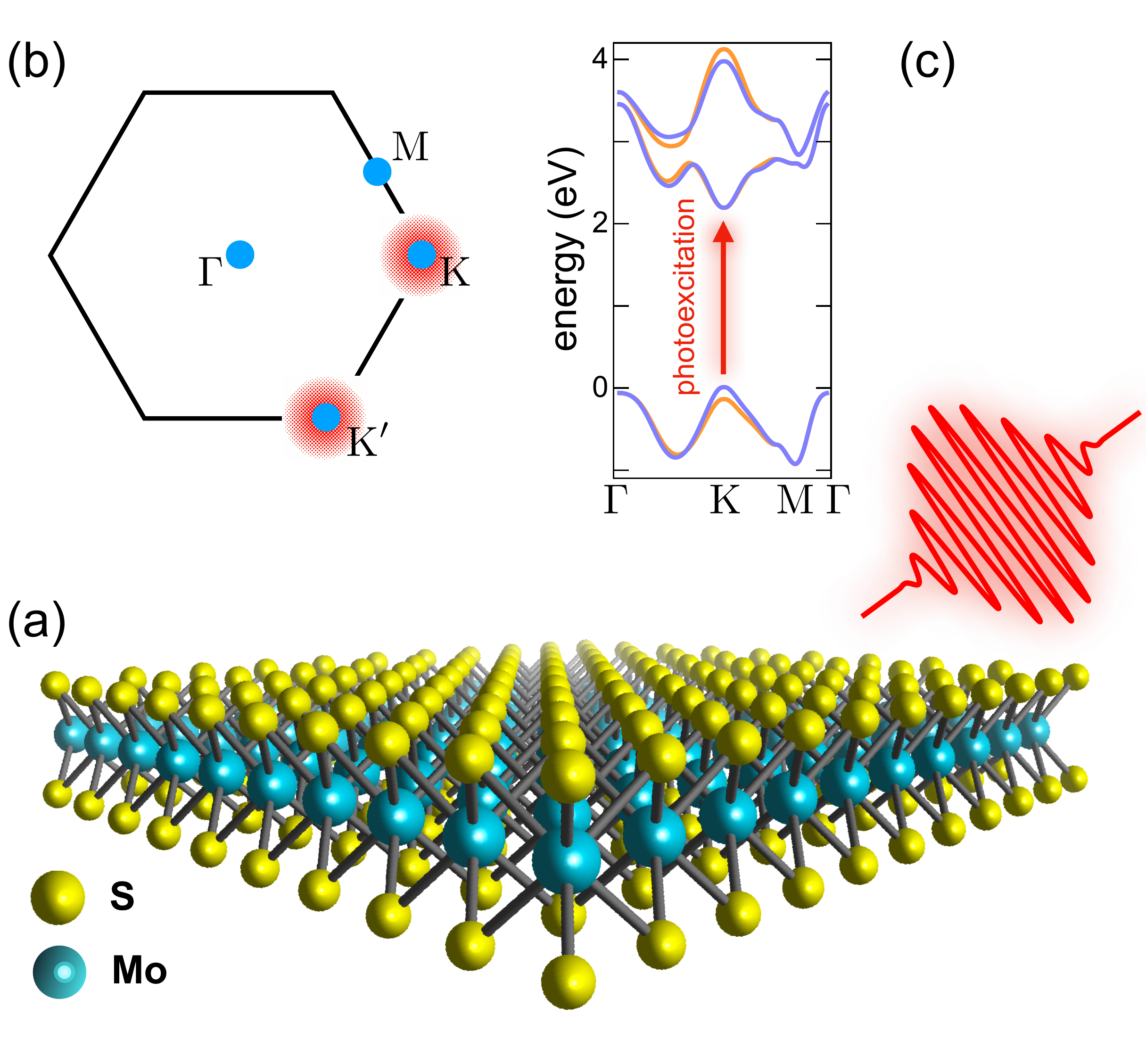}}
\caption{\footnotesize{\textbf{Coherent-to-incoherent crossover.} 
(a) Monolayer MoS$_{2}$  photoexcited by an ultrafast ($15$~fs) VIS 
($516$~nm) pulse. (b) First Brillouin zone and the high-symmetry points (blue 
circles). The VIS pulse excites electrons close to the $K$ point. 
(c) Band structure including spin-orbit interaction.
}}
 \label{fig1}
\end{figure*}

We here put NEGF at work in a molybdenum disulfide (MoS$_{2}$)
monolayer (Figure~\ref{fig1}a) pumped above gap. 
The direct band gap, tunable in the range $2.0\div 2.6$~eV by varying the 
substrate~\cite{PhysRevMaterials.2.084002}, and the
possibility of selectively excite one of the two degenerate 
valleys with circular polarized light~\cite{PhysRevLett.108.196802,valleytronics1} make the 
MoS$_{2}$ monolayer a promising candidate 
for optoelectronic~\cite{GUPTA2020108200,KRISHNAN2019274} and valleytronic~\cite{valleytronics2} 
applications.
Above-gap excitations are relevant for photochemical 
catalysis,   
photovoltaic and thermoelectric energy conversion 
due to the chemical-physical processes triggered by 
the excess energy of carriers.
Our investigation shows that the coherent-to-incoherent crossover 
is characterized by a dephasing time $\tau_{\rm deph}$  which depends 
 on the excitation density $n$ as
 $\tau_{\rm deph}^{-1}=\tau_{0}^{-1}+cn^{1/3}$. 
The lattice coherence survives much longer than the electronic 
coherence~\cite{PhysRevB.107.054102}. Several hundreds of 
femtoseconds after the crossover the 
laser-induced nuclear displacements still oscillate undamped.  
The dephasing process is accompanied by a highly complex dynamics of
carriers and phonons.
Intra-valley scattering -- mediated by 
both particle-hole and optical/acoustic phonon emission~\cite{acoustic,Asakura2021} -- is responsible for 
an ultrafast migration of hot electrons and holes toward the band 
edges already during pumping, giving rise to spots of different quasi-thermal 
carrier distributions shortly after the photoexcitation~\cite{nie2014ultrafast}. Inter-valley 
scattering is a slightly slower process, and it is pivotal for 
reaching a homogeneous thermal 
distribution across the entire Brillouin zone~\cite{caruso}. 
We can distinguish a fast and a slow timescales for the inter-valley 
scattering. 
Total energy conservation enables us to 
shed light on the tangled energy exchange between the electronic and 
phononic subsystems. We find that the energy gain in the build-up 
of carrier dressing by phonons is the main responsible for 
the increase of the lattice temperature during the first $50-60$~fs.
Thereafter the still hot dressed carriers begin to loose their excess 
energy at a rate of several hundreds of femtoseconds, causing a 
further increase of the lattice temperature until the end of the 
thermalization process.



\section{Photoexcited dynamics}

We study the photoexcited 
dynamics of a MoS$_{2}$ monolayer (Figure~\ref{fig1}a) 
initially at temperature $T=100$~K 
driven out of equilibrium by a linearly polarized 15~fs 
pulse with central frequency $\w_{0}$ exceeding the bandgap  by $0.2$~eV.  
In our simulations the pump fluence has been varied to span
a wide range of excitation densities  from $10^{10}$ up to
$10^{14}~\rm{cm}^{-2}$.
The excess energy  of the hot carriers is released
via competing and interconnected processes
mediated by both {\it el-el} and 
{\it el-ph} scattering, each one of them having their own timescale. 
The theoretical analysis of such a complex scenario cannot rely 
on simplified model Hamiltonians.
We therefore solve the NEGF equations of motion for the electronic 
and phononic density matrices $\r_{\blk ij}$ and $\g_{\blk \m\n}$ 
(all diagonal and off-diagonal elements are included) using  
spin-orbit dependent electronic bands $\epsilon_{\mathbf{k}i}$  
(Figure~\ref{fig1}b-c), 
with $i$ the spin-band index ($i=v,c$ for valence and conduction 
respectively) and $\mathbf{k}$ the two-dimensional crystal momentum, 
phononic bands $\omega_{\mathbf{k} \nu }$, 
with $\nu$ the phonon branch, Coulomb interaction
$V_{imjn}^{\blq\blk \blk'}$
and electron-phonon couplings
$g^{\nu}_{\blk i \blk' j}$
as input~\cite{PhysRevB.88.085433,PhysRevB.87.115418} (see Note 1 in the Supporting Information 
for details). 
The Coulomb integrals responsible for the formation of excitons have the form $V_{cvv'c'}$ and 
$V_{vcc'v'}$, and are explicitly included.

The energy-conserving nature of our approach and  the
simultaneous inclusion of the time non-local $GW$ and FM
self-energies allows for 
describing the multiscale dynamics 
as well as 
the ultrafast energy exchange between electrons and the underlying 
lattice.

\begin{figure*}[tbp]
\centerline{\includegraphics[width=.99\textwidth]{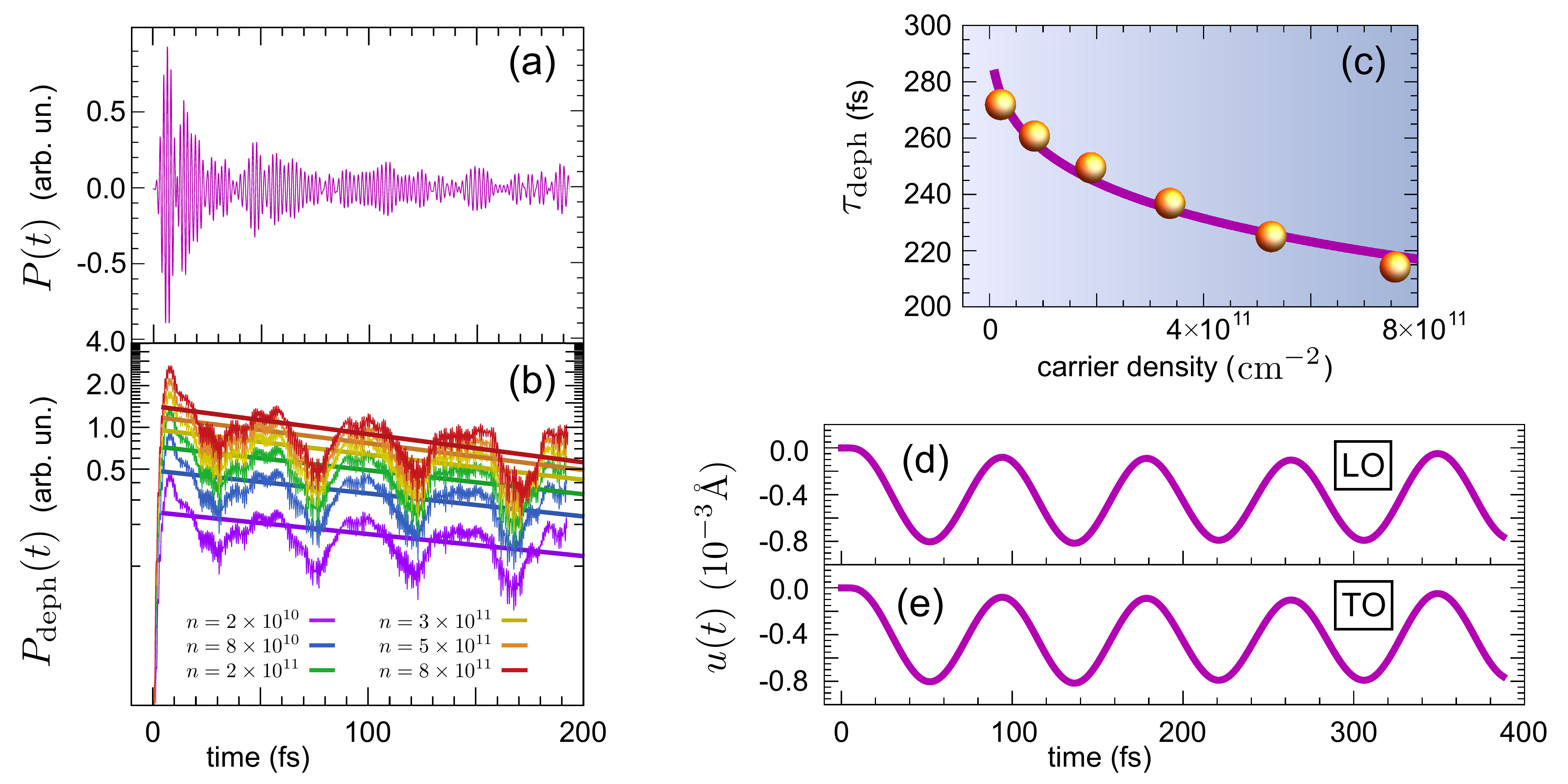}}
\caption{\footnotesize{\textbf{Coherent-to-incoherent crossover.} 
(a) Damping of 
the macroscopic polarization $P$ due to interference and dephasing 
for carrier density $n=3\times 10^{13}~\rm{cm}^{-2}$. 
(b) Logarithmic plot of $P_{\rm deph}$ for different excited carrier densities.
(c) Dependence of $\t_{\rm deph}$ on the excited carrier density (circles) and 
best fit with the function $\tau_{\rm 
deph}[n]=\frac{\tau_{0}}{1+\tau_{0}cn^{1/3}}$. 
(d-e) Time-dependent nuclear displacement of the LO
(top) and  TO (bottom) phonons for carrier density $n=3\times 10^{13}~\rm{cm}^{-2}$. Phonon coherence is preserved during 
the whole simulations window  ($\sim 400$~fs).
}}
 \label{fig2}
 \end{figure*}
 
\subsection{Correlation-induced dephasing}

The photoexcitation initially generates 
a hot electron-hole plasma around the $K$ and $K'$ 
points (Figure~\ref{fig1}b-c), more precisely at
the conduction  and valence 
 energies that
 approximatively match the condition 
$\epsilon_{\mathbf{k}c}-\epsilon_{\mathbf{k}v} \simeq \hbar \omega_{0}$.
The plasma inherits the coherence of the laser 
pulse, as demonstrated by the oscillating 
macroscopic polarization
displayed in Figure~\ref{fig2}a.
The finite life-time of the polarization is due to
interference and dephasing processes. 
The former are responsible for the so-called ``free-induction 
decay''~\cite{RevModPhys.74.895},
a phenomenon not related to correlations. 
As each optical transition  oscillates with its own 
frequency $\epsilon_{\mathbf{k}c}-\epsilon_{\mathbf{k}v}$, 
the oscillations get 
rapidly out of phase causing an ultrafast collapse of the 
macroscopic polarization. 
The dephasing, instead, destroys the coherence of the electron-hole 
pairs and convert them into an incoherent admixture of pairs. 
This process is governed by {\it 
el-el} and {\it el-ph} 
scattering, and it is distinct from the free-induction 
decay since all
$\mathbf{k}$-components $P_{\blk}(t)=\sum_{cv}\Re[ 
d_{vc\blk}\r_{\blk cv}]$, $d_{\blk cv}$ being the valence-conduction 
dipole matrix elements along the polarization direction of the laser
(see Note 1 in the Supporting Information), 
of the macroscopic polarization $P(t)=\sum_{\blk}P_{\blk}(t)$  
damp.
In order to assess the correlation-induced 
dephasing time $\tau_{\rm deph}$ we then introduce the 
more sensible quantity $P_{\rm deph}(t)=\sum_{\blk}|P_{\blk}(t)|$.
NEGF simulations in a two-band jellium-like model 
reported a dependence of 
$\tau_{\rm deph}$ on the excited carrier density 
$n$~\cite{PhysRevB.62.2686};
in both two-dimensional (2D) and three-dimensional
(3D) materials $\tau_{\rm deph}$ scales with a power low dependence 
on the carrier density $n$ according to
$\tau_{\rm deph}^{-1}=\tau_{0}^{-1}+cn^{1/3}$,
where $\tau_{0}$ is a density-independent contribution due to {\it 
el-ph} interactions while
the  power-law $n^{1/3}$ is an unambiguous signature of
the non-Markovian quantum kinetics ruled by the {\it el-el} 
interaction. 
This evidence has been experimentally confirmed in 2D and 3D GaAs through 
measurements of the time-integrated photon echo signal
for excitation densities 
$n\approx 10^{10}-10^{{12}}~\mathrm{cm}^{-2}$~\cite{PhysRevB.62.2686}.

In Figure~\ref{fig2}b we display the logarithmic plot 
of $P_{\rm deph}(t)$ 
for different pump fluences, leading to excited carrier densities in the range
$10^{10}-10^{{12}}~\mathrm{cm}^{-2}$. The  
fits with the exponential function $e^{-t/\tau_{\rm deph}}$ are shown as 
straight lines and demonstrate that the dephasing time 
decreases with increasing $n$.
The dependence of $\t_{\rm deph}$ on the 
excited carrier density is explicitly reported
in Figure~\ref{fig2}c. 
To verify whether the simulated dynamics
is compatible with the universal scenario of Ref.~\cite{PhysRevB.62.2686} we best fit 
our data  with the function 
$\tau_{\rm deph}[n]=\frac{\tau_{0}}{1+\tau_{0}cn^{1/3}}$,
obtaining the values $\tau_{0}=311$~fs and $c=1.50\times 
10^{-7}\mathrm{fs}^{-1}\mathrm{cm}^{2/3}$.
The agreement between the best fit  (solid line)
and the simulation data-points 
indicates that the dephasing 
is indeed driven by a non-Markovian dynamics.
For larger excitation densities we find that a single exponential is 
not sufficient to fit $P_{\rm deph}(t)$ and that a double exponential 
is more appropriate (Figure S2). This agrees with experimental  
findings in a five-layer MoS$_{2}$~\cite{nie2014ultrafast}. 

The inclusion of the Ehrenfest self-energy gives us  
access to the time-dependent nuclear displacements $u_{\n}(t)$
along the different normal modes.
In Figure~\ref{fig2}d-e we plot $u_{\n}(t)$ for the two most 
coupled optical phonons, i.e., the transverse optical (TO) and 
longitudinal optical (LO) phonons.
Due to the finite carrier density in the conduction band
the nuclei oscillate
around a non-equilibrium position with a period $T_{\n}$
of about $85$~fs, in agreement with the corresponding phonon 
frequency $\w_{{\mathbf 0}\n}=2\p/T_{\n}$.
Contrary to the electronic case the 
nuclear coherence is very long lived,
as no damping is observed during the whole  
simulation time (about $400$~fs).
This is consistent with a recent experimental work on a MoS$_{2}$
monolayer photoexcited by a ultrashort ($<20$~fs) pulse
reporting that phonon coherence survives for several picoseconds 
after the photoexcitation~\cite{trovatello}.

\begin{figure*}[tbp]
\centerline{\includegraphics[width=.8\textwidth]{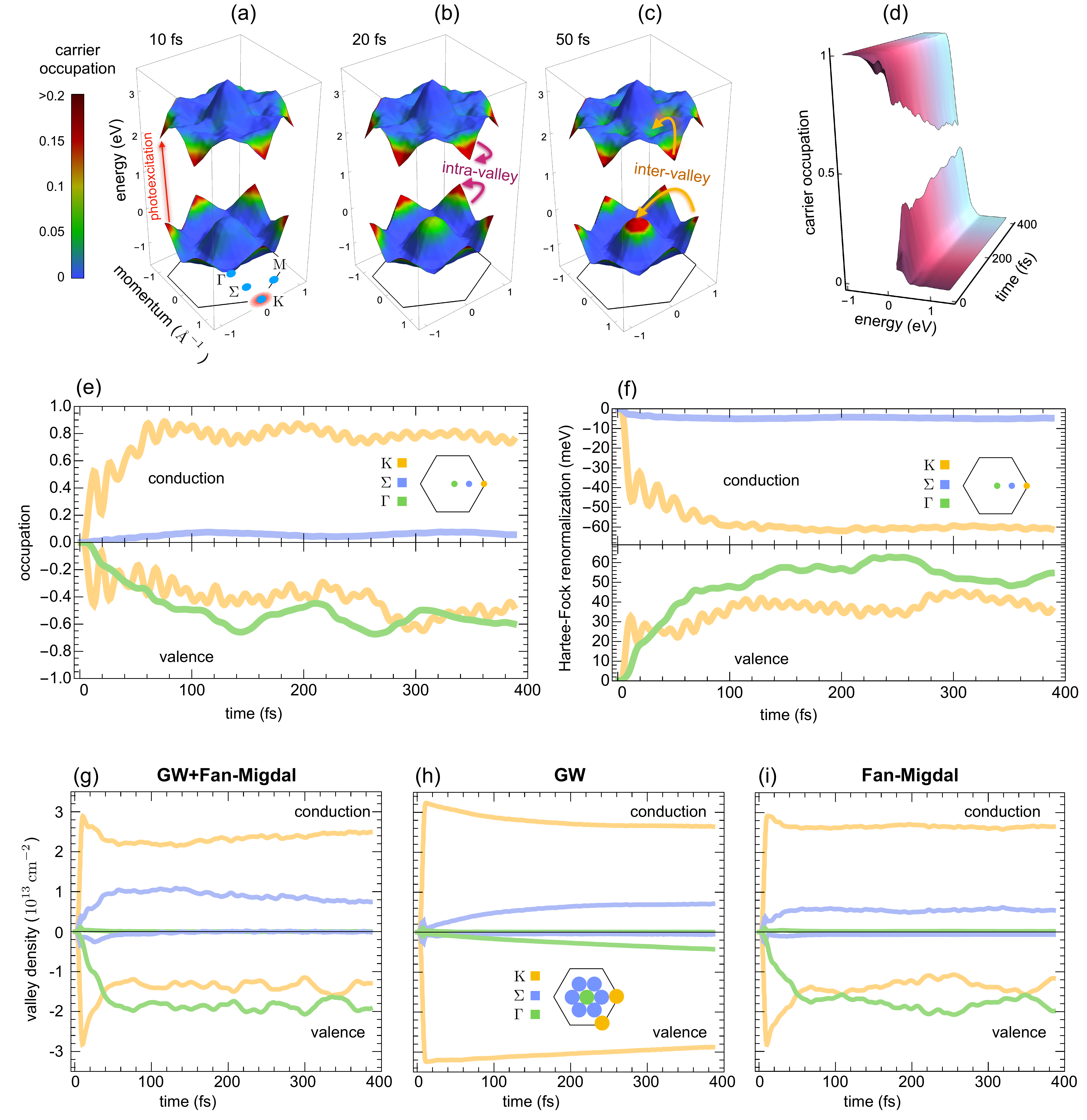}}
\caption{\footnotesize{\textbf{Carrier dynamics} 
(a-c) Snapshots of the $\blk$-resolved electron (for conduction 
bands) and hole (for valence bands) populations in the first 
Brillouing zone at times $t=10$~fs, $20$~fs and $30$~fs for a 
MoS$_{2}$ monolayer driven out of equilibrium by a laser pump of 
duration $15$~fs and central frequency $0.2$~eV larger than the 
direct gap at the $K$ and $K'$ points. (d) Energy dependent 
hole populations (for negative energies) and electron populations 
(for positive energies) as defined in the main text versus time.
(e) Time-dependent population at three high-symmetry points 
of the first Brilluin zone: population
of the conduction $K$ point $n_{Kc}(t)$ and $\S$ point $n_{\S c}(t)$, 
and valence $K$ point $n_{Kv}(t)-2$ and $\G$ point $n_{\G v}(t)-2$.
(f) Time-dependent modification of Hartree-Fock energies
at the same high-symmetry points as in panel (e). 
(g-i) Electronic density in the $K$ and $\S$ valleys of the conduction 
band (top) and hole density in the $K$ and $\G$ valleys of the valence 
band (bottom) versus time as obtained using NEGF (see Notes 2-3 in the Supporting Information)
in the $GW$+FM, $GW$-only  and FM-only 
approximations. In all panels the  carrier density is $n=3\times 
10^{13}~\rm{cm}^{-2}$.   }}
 \label{fig3}
 \end{figure*}

\subsection{Carrier relaxation}

The dynamics of electrons and holes  initially photoexcited around the $K$ 
and $K'$ valleys is extremely rich and the effects of {\it el-el} and 
{\it el-ph} interactions is not simply additive.
In Figure~\ref{fig3}a-c we provide a comprehensive picture of the energy-momentum resolved
carrier dynamics for a relatively high excitation density $n=3\times 
10^{{13}}~\mathrm{cm}^{-2}$ during the first $50$~fs.
Intra-valley scattering and the build-up of the Coulomb 
screening 
are the fastest processes taking place~\cite{PhysRevB.62.7179},
influencing the dynamics already  during  illumination. The intrinsic 
anisotropy introduced by the linearly polarized  
pulse (Figure S3) is almost entirely washed out already at time 
$t=10$~fs (Figure~\ref{fig3}a).

In Figure~\ref{fig3}d we show the energy-dependent occupations of 
electrons $n_{e}(\e)=\sum_{(c,\blk):\e_{c,\blk}=\e}n_{c,\blk}$ 
and holes $n_{h}(\e)=\sum_{(v,\blk):\e_{v,\blk}=\e}(1-n_{v,\blk})$, where the 
sums run over all $\blk$ and over all conduction and valence bands, 
and the energy is measured with respect to the 
conduction band minimum and valence 
band maximum respectively. Right after pumping the carrier distribution 
is highly non-thermal, with peaks at the energy 
satisfying the resonant condition 
$\epsilon_{\mathbf{k}c}-\epsilon_{\mathbf{k}v} \approx 
\hbar \omega_{0}$. Although the energy-dependent occupations
reach a Fermi-Dirac distribution  
within the first 40-50~fs the electronic subsystem has not yet thermalized 
since degenerate 
$\blk$-points in the first Brillouin zone are not equally populated, 
see for instance the $K$ and $\G$ points in the valence band in 
Figure~\ref{fig3}e. 
The thermalization takes longer than the dephasing as it is still not 
reached at the end of our simulation time. 
The two mechanisms contributing to the 
thermalization are intra- and inter-valley scattering.

\begin{figure*}[tbp]
\centerline{\includegraphics[width=.99\textwidth]{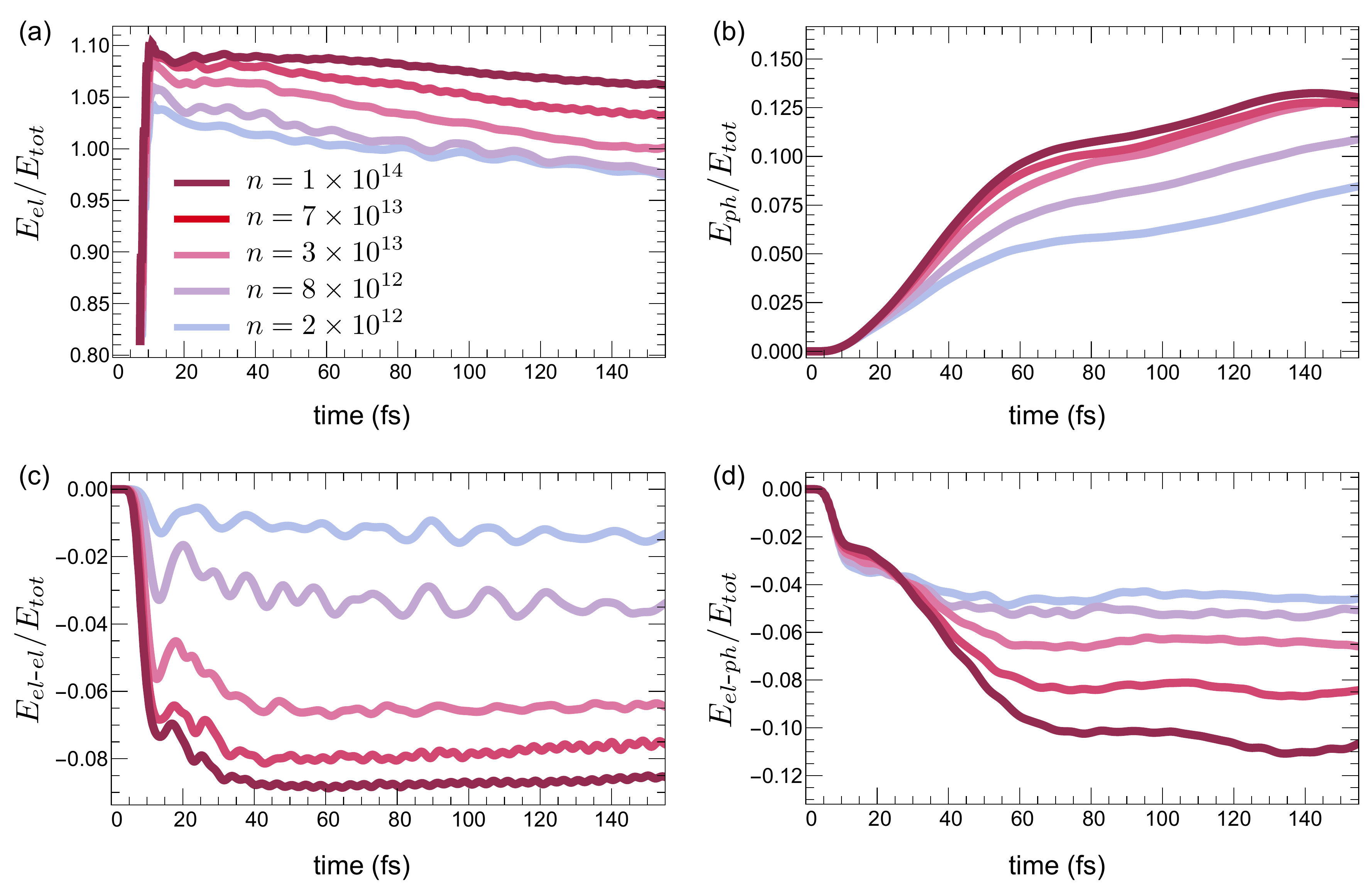}}
\caption{\footnotesize{\textbf{Energy redistribution} 
Electronic energy $E_{el}$ [measured with respect to its equilibrium value 
$E_{el}(t=0)$] (a), phononic energy $E_{ph}$ (b), 
$el$-$el$ correlation energy $E_{el\mbox{-}el}$ (c) and 
$el$-$ph$ correlation energy $E_{el\mbox{-}ph}$ (d) renormalized to 
the total energy $E_{tot}$ for different densities of the excited 
carriers.}}
\label{fig4}
\end{figure*}

{\bf Intra-valley scattering -- }
The intra-valley scattering  is 
responsible for the 
relaxation of carriers toward the bottom of the band valleys 
(Figure~\ref{fig3}e). We can distinguish two distinct time-scales for the 
initially photoexcited $K$ points. 
While the laser pulse is active electrons predominantly lose their 
excess energy due to inelastic $el$-$el$ collisions, thereby 
causing an ultrafast 
increase of the electronic correlation energy, see Figure~\ref{fig4}a 
and \ref{fig4}c. After the photoexcitation intra-valley scattering 
mainly occurs via  
the emission of optical and acoustic 
phonons around the $\Gamma$ point~\cite{acoustic,Asakura2021}, 
the optical contribution being more important 
than the acoustic one~\cite{PhysRevResearch.3.023072}.
The time-scale for reaching a 
quasi-steady value of the occupations at the $K$ points
is about 
$60$~fs (Figure~\ref{fig3}e) which is the same time-scale 
for the initial raise of the phonon energy 
(Figure~\ref{fig4}b).
Interestingly, in this time-window the electronic energy is (roughly) 
constant up to $\sim 40$~fs  
(Figure~\ref{fig4}a). In fact, the migration toward the bottom of the 
valley is accompanied by a renormalization of the one-particle 
energies (Figure~\ref{fig3}f). The emission of phonons in this 
time-window is made possible by the energy gained in dressing 
electrons and holes with phonons. 
Such ultrafast process can only be 
captured  with an energy-conserving approach.
Inspection of Figure~\ref{fig4}b and 
~\ref{fig4}d shows that the rate at which 
the phonon energy increases is the same as the rate 
at which the electron-phonon correlation energy (an indicator of phononic 
dressing) decreases.
We also report that holes relax faster than electrons 
(see flank at positive energy in Figure~\ref{fig3}d), 
in qualitative agreeement with
recent BE-based simulations~\cite{caruso}.
This is due to the transient trapping of electrons at the bottom of 
the $\Sigma$ valley, located about $0.2$~eV above the bottom of 
the $K$ valley.

The electronic correlation energy (Figure~\ref{fig4}c)
is also an indicator of the 
{\em build-up of screening} by the electron-hole plasma. 
After the pump ($15$~fs) the correlation energy first decreases 
(despite the superimposed oscillations) on a 
time-scale of $40$~fs and then begins to increase on a much longer 
time-scale. This second slower stage of the dynamics is mainly 
characterized by the equilibration of the populations in different 
valleys.

{\bf Inter-valley scattering -- } 
Phonon emission and the build up of screening are concomitant 
with the second mechanism, i.e., inter-valley scattering. 
In the first $60$~fs an important fraction of excited electrons
migrate toward the six degenerate $\Sigma$ valleys~\cite{bandgap1}
and an even larger amount of holes migrate towards the $\Gamma$
valley (Figure~\ref{fig3}c and ~\ref{fig3}g)~\cite{xu2021}. 
In this time-window
inter-valley scattering is mainly a phonon-mediated process, and 
zone-edge phonons (both optical and acoustic) play a crucial 
role~\cite{edgeph1,D0NR04761A,edgeph,edge2}. 
This is confirmed by our simulations (see below).
Another evidence in favor of the phonon-mediated mechanism is 
provided in Figure~\ref{fig3}i where we include only $el$-$ph$ 
scattering through the FM self-energy, 
and still observe the $60$~fs time-scale. Comparing the FM-only 
dynamics with the $GW$-only dynamics (Figure~\ref{fig3}h) we infer 
that $el$-$el$ scattering is much slower in moving electrons between 
different valleys, although equally important in terms of 
effectiveness. Inter-valley scattering in $GW$-only occurs through 
particle-hole exchange which is initially suppressed by momentum 
conservation. The full $GW$+FM dynamics (Figure~\ref{fig3}g) 
reveals an overshooting of charge at the $\S$ valley of the 
conduction bands and a rebound of charge between the almost degenerate 
$K$ and $\G$ valleys in the valence bands. The near degeneracy of 
the $K$ and $\G$ valleys are the cause of the long thermalization 
time in the MoS$_{2}$ monolayer. 

Although phonon-exchange and particle-hole exchange live on different 
time-scales they cannot be treated 
independently. In fact, the $GW$+FM dynamics is not merely the 
FM-only dynamics for times $t<60$~fs and the $GW$-only dynamics for 
$t>60$~fs (Figure~\ref{fig3}e-g). 
The role of particle-hole exchange diminishes with decreasing the 
pump fluence. It is only in the regime of weak pumps that the 
inter-valley FM-only and $GW$+FM dynamics become similar (not shown). 

The four contributions to 
the total energy (electronic, phononic, electron-electron 
 and electron-phonon) 
are shown in Figure~\ref{fig4}a-d for different 
densities of the excited carriers. The total energy is correctly 
constant after the pump (times $t>15$~fs, not shown). 
The electronic energy $E_{el}$ remains about ten times larger 
than the phononic 
energy $E_{ph}$ whereas the $el$-$el$ and $el$-$ph$ correlation 
energies, $E_{el\mbox{-}el}$ and $E_{el\mbox{-}ph}$,  are 
comparable in size. We observe that $E_{el\mbox{-}ph}$
saturates after $60$~fs (this is the time-scale of the 
phonon mediated inter-valley scattering) while all other contributions 
show an almost linear behavior from times $t\sim 60$~fs up to hundreds of 
femtoseconds (this is the time-scale of the 
particle-hole mediated inter-valley scattering).

\subsection{Non-equilibrium phonons}

The ultrafast photoexcitation of carries
promptly activates the dynamics of  phonons,
whose full density matrix $\g_{\blk \m\n}$ 
is evolved in time along with the electronic density matrix 
$\r_{\blk ij}$ and the nuclear displacement 
$u_{\n}$ and momentum $p_{\n}$ of the $\n$-th mode (see Notes 2-3 in the Supporting Information 
for details).
Through the simultaneous
propagation of all these quantities the electronic 
(phononic) feedback on the phononic (electronic) subsystem is 
properly accounted for, thereby guaranteeing
the conservation of the total energy.
The NEGF approach gives us access 
to several relevant quantities like
the mode- and momentum-resolved phonon
populations $n_{ 
\mathbf{k}\nu}=\frac{1}{2}(\Tr[\g_{\blk \n\n}]-1)+\frac{1}{2}
\d_{\blk,0}(u^{2}_{\n}+p^{2}_{\n})$, 
the mean-squared nuclear displacement $\bra 
u^{2}_{\n}\ket=[\g_{\bf{0}\n\n }]_{11}$ and momentum 
$\bra p^{2}_{\n}\ket=[\g_{\bf{0}\n\n }]_{22}$ and the  {\it el-ph} correlation energy.
The effective  temperature of the different phonon modes 
has been estimated according to
$T_{ \mathbf{k} \nu}=\hbar \omega_{\mathbf{k} \nu } [k_{\mathrm{B}} \ln 
(1+\frac1{n_{ \mathbf{k}\nu}})
]^{-1}$, and the effective (average) lattice temperature according to 
$T_{av}=\frac{1}{N_{\mathbf{k}}N_{ph}}\sum_{\n\mathbf{k}}
T_{ \mathbf{k}\nu}$,
where $N_{ph}=9$ is the total number of phonon branches and 
$N_{\mathbf{k}}$ is the number of discretized $\mathbf{k}$-points. 

  \begin{figure*}[tbp]
\centerline{\includegraphics[width=.99\textwidth]{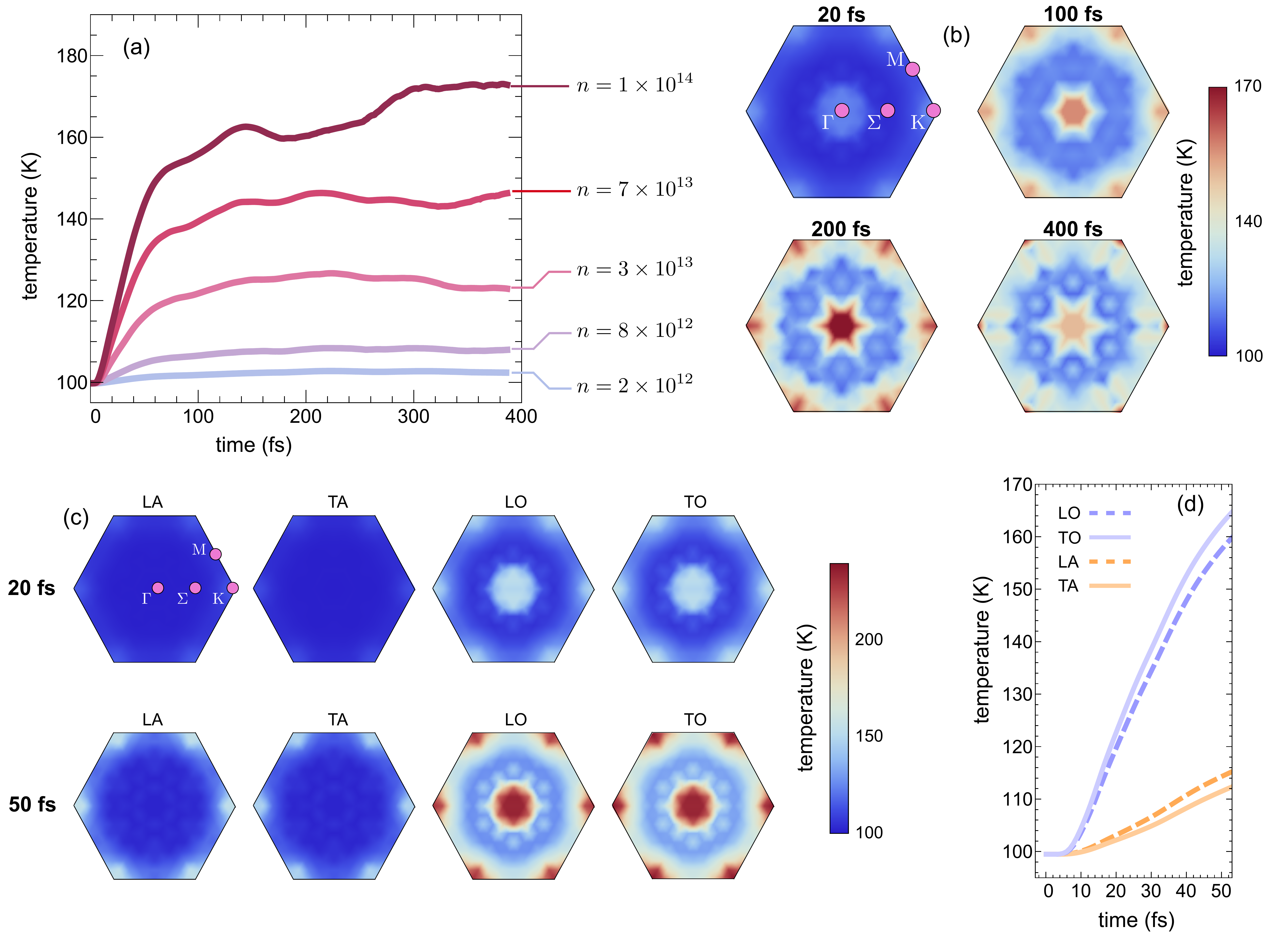}}
\caption{\footnotesize{\textbf{Phonon Dynamics.}
(a) Average temperature versus 
time for different densities expressed in $\rm{cm}^{-1}$.
(b) Average temperature map 
from time $t=20$~fs to time $400$~fs.
(c) 
Branch-resolved temperature maps 
at times $t=20$~fs and $50$~fs.  
(d) Branch-resolved temperature for the LA, TA, LO and TO 
phonons versus time.
In panels (b-d) the carrier density is $n=3\times 10^{13}~\rm{cm}^{-2}$.
} 
}
 \label{fig5}
 \end{figure*}

In all our simulations the phonons are initially
in thermal equilibrium at a lattice temperature $T=100$~K.
For a wide range of excitation densities the excess energy of hot 
carriers is transferred to the lattice within the 
first few hundreds of femtoseconds (Figure~\ref{fig5}a), in agreement 
with recent experiments~\cite{phonoheat} and  
simulations~\cite{phonoheat2}.
For moderate and high excitation density the energy transfer is 
characterized by a rapid (within the first $60$~fs)
increase followed by a 
slower increase of $T_{av}$.
The final average temperature of 175~K predicted for $n=10^{{14}}~\mathrm{cm}^{-2}$ 
is in excellent agreement with recent BE-based simulations in 
the same material at the same excitation density~\cite{caruso}.
Although $T_{av}$ is fairly constant after a few hundreds 
of femtoseconds the $\blk$-dependent temperature remains highly 
inhomogeneous at this time-scale (Figure~\ref{fig5}b).
In fact, phonons loose coherence and eventually thermalize on a 
much longer time-scale than 
electrons, see the undamped oscillations 
of the nuclear displacements in the first $400$-fs 
reported in Figure~\ref{fig2}d-e.
The energy and momentum constraints of the
phonon-assisted energy loss limit the emission of phonons 
to a few momenta in the first Brillouin zone. We have singled out 
the emission of optical phonons around the $\Gamma$ point as a  
mechanism for intra-valley scattering and the emission 
of  zone-edge acoustic and 
optical phonons as a mechanism for inter-valley 
scattering~\cite{edgeph1,D0NR04761A,edgeph,edge2}.
It follows that lattice heating is confined around a few 
high-symmetry points~\cite{caruso} (Figure~\ref{fig5}b), in agreeement with
recent experiments in MoS$_{2}$~\cite{caruso2}.
This is a general feature of materials hosting several valleys.
Simplified approaches like the two-temperature 
model are highly questionable in this context, at least during 
the first few picoseconds. 
In Figure~\ref{fig5}c we show the branch-resolved 
temperature map for the longitudial acoustic (LA), transverse 
acoustic (TA), LO and TO phonons at times $t=20$~fs 
and $50$~fs. 
Acoustic phonons are mainly emitted at the $\G$ point and contribute 
for about $20\%$ to the overall heating. Optical phonons are mainly emitted 
at the $\G$, $K$, $K'$ and to a lesser extent at the $\S$ points, 
in agreement with the 
fact that the electronic density in one of the $\S$ valleys 
(accompanied by the emission of $\S$-phonons) is a 
factor of ten smaller than the hole density at $\G$ 
(accompanied by the emission of $K$ and $K'$-phonons). 
The initial raise of the 
temperature $T_{\n}$ of the four phonon branches versus 
time is reported in Figure~\ref{fig5}d. We observe a substantial 
balance between the two optical and the two acoustic modes.

\section{Conclusions and outlooks}

Atomically thin two-dimensional materials like graphene and the whole 
family of TMD have broken into the research topics of modern 
science due to their desirable optoelectronic and valleytronic properties for 
nanoelectronics, photonics and quantum computing,  
as well as for their high surface-to-volume ratio for 
photovoltaic and sensing applications. Time-resolved spectroscopies 
offer an unprecedented tool to characterize these materials and 
to validate theoretical concepts and techniques indispensable for
rationalizing and guiding future research. Developing such techniques 
is a major challenge due to the simultaneous interplay of 
light-matter, $el$-$el$ and $el$-$ph$ interactions in an out-of-equilibrium context.

In this work we have presented an advanced ab initio many-body method 
for real time simulations of 2D systems which encompasses 
several fundamental aspects, namely 
the non-Markovian nature of the dynamics for both carrier and 
polarization degrees of freedom, the  
electronic feedback on the nuclear degrees of freedom, 
the effects of nonequilibrium dynamical screening and the 
conservation of the total energy. 

The dynamics of a MoS$_{2}$ monolayer pumped above gap has been 
followed from the instant in which the laser impacts the material 
until the  coherent-to-incoherent crossover and beyond. 
Agreement with available experimental and theoretical findings
on specific aspects of the dynamics has been found.
Real-time simulations with/withouth $el$-$el$ and/or $el$-$ph$ 
correlations as well as comparisons with Markovian dynamics
allows for elucidating the role and the time-scales of 
entangled physical mechanisms. 
We point out that many critical
processes like the build-up of screening, carrier dressing by phonons,
transient lattice heating and intra- and inter-valley scattering
occur {\it during} the coherent-to-incoherent crossover.
The resulting scenario is therefore intrinsically complex and 
hardly interpretable in terms 
of independent processes.

We foresee a number of further studies and developments. 
The behavior of different 2D 
materials optically excited by laser pulses of varying 
duration, polarization and central frequencies can be 
characterized through 
band-, spin- and momentum-resolved carrier populations, 
polarization, branch- and momentum-resolved phonon populations, 
electron-electron, electron-hole or exciton, and electron-phonon or 
polaron correlation functions, nuclear displacements as well as 
kinetic, potential and correlation energy contributions.
The NEGF method can be systematically improved or adapted to the physical 
situation of interest. At high pump fluences the dynamical screening 
of the electron-phonon coupling may become relevant and it 
can be incorporated through the so called doubly-screened 
$G\tilde{W}$ approximation~\cite{GWtilde} without affecting the time-scaling. 
Nonlinear phonon coupling as well as phonon-phonon interactions can 
be included through the addition of properly chosen 
self-energy diagrams while preserving the energy conservation law.
We add that the time-linear NEGF scheme allows  easy parallelisation 
for high-throughput 
simulations on denser grids and/or multi-layer systems.
We envisage the implementation of the proposed method for 
parameter-free predictions in (quasi) 2D 
nonequilibrium materials.

%
%
%
%
%
%
%

\section{Acknowledgements}

The Authors acknowledge funding from MIUR PRIN Grant
No. 20173B72NB, from INFN17-Nemesys project, 
Tor Vergata University for financial
support through Projects 2DUTOPI and TESLA.

%
%
%
%
%
%
%
%


\providecommand{\latin}[1]{#1}
\makeatletter
\providecommand{\doi}
  {\begingroup\let\do\@makeother\dospecials
  \catcode`\{=1 \catcode`\}=2 \doi@aux}
\providecommand{\doi@aux}[1]{\endgroup\texttt{#1}}
\makeatother
\providecommand*\mcitethebibliography{\thebibliography}
\csname @ifundefined\endcsname{endmcitethebibliography}
  {\let\endmcitethebibliography\endthebibliography}{}

\end{document}

%% file: ourmacros.tex
\newcommand{\be}{\begin{equation}}
\newcommand{\ee}{\end{equation}}
\newcommand{\bea}{\begin{eqnarray}}
\newcommand{\eea}{\end{eqnarray}}

\def\g{\gamma}
\def\G{\Gamma}
\def\d{\delta}

\def\e{\epsilon}

\def\m{\mu}
\def\n{\nu}

\def\p{\pi}

\def\r{\rho}

\def\S{\Sigma}
\def\t{\tau}

\def\w{\omega}

\def\blk{{\mathbf k}}

\def\blq{{\mathbf q}}

\def\bra{\langle}
\def\ket{\rangle}

\def\Tr{{\rm Tr}}
\def\Re{{\rm Re}}

\def\1op{\hat{\mathbbm{1}}}